\documentclass[11pt]{article}

\usepackage{graphicx}
\usepackage{amsmath}
\usepackage{amsfonts}
\usepackage{amssymb}
\usepackage{color}
\usepackage{amscd}
\usepackage{bm}
\usepackage{dcolumn}
\usepackage{float}

\usepackage[latin1]{inputenc}
\usepackage[spanish,english]{babel}
 \usepackage{setspace}
 \usepackage{verbatim}
\usepackage{soul}

\usepackage[margin=2.5cm]{geometry}

\begin{document}

\begin{titlepage}

\begin{center}
{\bf The holographic state of matter and its implications for quantum gravity}
\end{center}

\begin{center}
Yong Xiao$^{\dag*}$ \let\thefootnote\relax\footnotetext{$^*$Corresponding author, xiaoyong@hbu.edu.cn} \\
{\small \it {$\dag$ Key Laboratory of High-precision Computation and Application of Quantum Field Theory of Hebei Province,
College of Physical Science and Technology, Hebei University, Baoding 071002, China}}
\end{center}

\begin{abstract}
The holographic state of matter exists in the quantum gravitational regime, with black holes as the example. In this essay, we provide a microscopic derivation to the holographic thermodynamics and further discuss its implications for quantum gravity. Especially, we establish the link between holographic entropy and dimensional reduction. It seems that the fundamental physics behind black holes and the very early universe is $1+1$ dimensional.
\end{abstract}

\vspace{3.6cm}

  \centerline{\sffamily Essay
written for the Gravity Research Foundation 2021 Awards for Essays
on Gravitation}\vspace{0.25cm}
 \centerline{\sffamily Submitted on March 29, 2021 }

\end{titlepage}
\uppercase\expandafter{\romannumeral1}. \textbf{Introduciton}
\vspace{0.12cm}

In the early days, only several simple states of matter had been recognized: gas, liquid, solid and plasma. However, nowadays many new quantum states of matter have been found in extreme physical environments, for instance, Bose--Einstein condensates at ultra-cold temperatures. These states of matter can transition among each other smoothly or abruptly according to the circumstances.

 There exists a holographic state of matter in the ultraviolet regime where the quantum gravitational interaction dominates the system, with black holes as the example. To be concrete, one starts from a natural question: for a given energy $E$ confined in a given volume $V$, what is the maximum entropy $S(E,V)$? An answer would be $S\sim E^{\frac{3}{4}}V^{\frac{1}{4}}$, which is that for the relativistic particles such as photons, from the knowledge of ordinary quantum field theory (QFT). But the ordinary QFT may become invalid at the extremely high energetic and small length scales. In fact, many approaches have pointed to a new form \cite{Sasakura:1999xp,Brustein:1999md,mathur}
\begin{align}
S\sim \sqrt{\frac{EV}{G}}.  \label{start}
\end{align}
The presence of $G=l_p^2$ has illustrated that the formula may play an important role in the quantum gravitational regime, especially in the study of black hole physics and the very early universe.  We can call this new state of matter satisfying eq.\eqref{start} as ``holographic'' because it can readily explain the area entropy. Obviously one gets the holographic entropy $S\sim A/G$ with $E\sim R/G, V\sim R^3$. Besides, the formula \eqref{start} has many intriguing properties.  For example, the formula is invariant under the T and S dualities, and it even keeps its form when one starts from a higher dimensional space-time and curls up some extra dimensions, since $\sqrt{\frac{E V_3 (L_c)^{D-4}}{G_{D}}}=\sqrt{\frac{E V_3}{G}}$ where $(L_c)^{D-4}$ is the volume of the extra dimensions \cite{mathur}.

In this essay, we provide a microscopic derivation to the formula \eqref{start} and the entire holographic thermodynamics, which reveals interesting implications for quantum gravity.

\vspace{0.37cm}

\uppercase\expandafter{\romannumeral2}. \textbf{The microscopic derivation to the holographic thermodynamics}
\vspace{0.12cm}

In statistical mechanics, we need a partition function to derive the thermodynamic behaviors of a system. The logarithm of the partition function for bosonic particles is
\begin{align}
\begin{split}
\ln \Xi = - g_1 \int_0^\infty  \ln ( 1 - e^{ - \beta \varepsilon }  ) D(\varepsilon) d\varepsilon,
 \end{split} \label{lnx}
\end{align}
where $\beta=1/T$, $D(\varepsilon)d\varepsilon$ is the number of single-particle quantum states with energy between $\varepsilon$ and $\varepsilon+d\varepsilon$, and $g_1$ represents other possible degrees of freedom such as polarization.

For ordinary QFT particles, according to the quantum principle $\triangle q_i \triangle p_i \geq \frac{\hbar}{2}$ and the relativistic energy-momentum relation $\varepsilon=c p$, the state density is found to be $D(\varepsilon) d\varepsilon  \sim V \varepsilon ^2 d\varepsilon$. Then it follows from eq.\eqref{lnx} the familiar photon-gas behaviors $E \sim  V T^4, S\sim V T^3$, and $w=\frac{P}{\rho}=\frac{1}{3}$. One can observe the relation $S\sim E^{\frac{3}{4}}V^{\frac{1}{4}}$.

In the quantum gravitational regime, with the dimensionful constant $G$ at hand, we may conjecture a new state density $g_1 D(\varepsilon) d\varepsilon=\frac{ 9 V}{\pi G }d\varepsilon$ \footnote{Its rough form is fixed by dimensional analysis, and the coefficient is chosen to recover the exact coefficients of black hole temperature and entropy. Actually, there is only one adjustable parameter $g_1$ and the Smarr formula $M=2 TS$ is independent of the choice of this parameter. }. Though the underlying microscopic principles are still unknown, this form is fixed by simple dimensional analysis. Then from eq.\eqref{lnx} we find \cite{Xiao:2019pkr,Xiao:2020hdt}
 \begin{align}
 E=\frac{3 \pi}{2 G}VT^2,\ \ \  S=\frac{3\pi}{G} VT \ \  \text{and}\ \ w=\frac{P}{\rho}=1, \label{holother}
 \end{align}
where we have set $\hbar=c=k_B=1$. Then the relation $S=\sqrt{6\pi} \sqrt{\frac{ EV}{G}}$ can be observed.

\vspace{0.37cm}

\uppercase\expandafter{\romannumeral3}. \textbf{The implications for quantum gravity}
\vspace{0.12cm}

For the holographic state of matter, we have derived the formula \eqref{start} and furthermore all the thermodynamic behaviors \eqref{holother}. Next we discuss its interesting implications for our understandings of quantum gravity as below. To support our viewpoints, we will also collect some evidences from the recent progress in the literature of quantum gravity.

\vspace{0.12cm}

1. \emph{The quantum gravitational systems are effectively $1+1$ dimensional. }
\vspace{0.1cm}

By staring at the thermodynamic behavior \eqref{holother}, we may soon realize that it resembles that of an ordinary $1+1$ QFT system:
 \begin{align}
 E\sim L_s T^2,\ \ \  S\sim  L_s T \ \  \text{and}\ \ w=1. \label{sss}
 \end{align}
 The resemblance is clear if we use $L_{s}\equiv \frac{V}{G}$ in eq.\eqref{holother}.  The form of the entropy $S  \sim \frac{1}{ \hbar c } L_{s} T$ signifies that we were studying as if a $1+1$-dimensional quantum system other than a $3+1$-dimensional quantum system. Notice the result still holds when generalized to $D=d+1$ dimensional cases. In the quantum gravitational regime, it seems that the microscopic particles are always quantized as if the space-time were $1+1$ dimensional, irrespective of the realistic space-time dimensions that they live.

 Our observation, though seemingly weird at first sight, is actually consistent with the recent progress in quantum gravity. Cumulative evidences have indicated that quantum gravity should be effectively $1+1$ dimensional at small sizes and recovers to be $D=d+1$ dimensional at large scales, from the analysis of Euclidean quantum gravity \cite{laus}, causal dynamical triangulation \cite{amb} and Ho\v{r}ava-Lifshitz Gravity \cite{horava}. For a review of the phenomenon of short-distance dimensional reduction, see \cite{carlip1,carlip2}. Ho\v{r}ava stressed that the behavior implies the existence of some special properties of quantum gravity at short distances, but not necessarily a foamy structure of space-time.
\vspace{0.12cm}

2. \emph{Black holes may resemble a ball of wool thread.}
\vspace{0.1cm}

We find that the black hole thermodynamics can be successfully explained from eq.\eqref{holother}.
 Taking the gravitational energy $M=(1+3w)E$ to be the energy of a black hole $\frac{R}{2G}$, we can easily deduce the exact Hawing temperature $T=\frac{1}{4\pi R}$ and the Bekentstein--Hawking entropy $S=\frac{A}{4G}$, as well as the Smarr formula $M=2TS$ for the Schwarzschild black hole.

 Generalizing the analysis to the higher dimensional black holes, we can derive the exact Smarr formula $M=\frac{D-2}{D-3}TS$ in $D=d+1$ dimensions, which is more than could be excepted  \cite{Xiao:2019pkr,Xiao:2020hdt}. Intriguingly, combined with $dM=TdS$, it gives $ d \ln S= \frac{D-2}{D-3}d \ln M$. This relates the exponent of the holographic entropy $S\sim M^\frac{D-2}{D-3}$ and coefficient of the Smarr relation.

The success in describing the black hole thermodynamics from eq.\eqref{holother}, or equivalently \eqref{sss}, provokes us to argue that black holes may be effectively a $1+1$ dimensional object. In this sense, a black hole may be appropriately doodled as a ball of wool thread.

We notice that Bekenstein and Mayo \cite{bema} suggested decades ago that black holes might be effectively $1+1$ dimensional by analyzing the entropy flow rate $\dot{S}$ through the horizon, which provides an independent supportive evidence to our argument. In addition, there were also other stringy scenarios for black holes from various motivations \cite{pol2,vene,Brustein}.
\vspace{0.12cm}

3.\emph{The very early universe is holographic and effectively $1+1$ dimensional.}
\vspace{0.1cm}

The high-energetic stage of the very early universe should be controlled by quantum gravity. Thus, from our analysis above, it is natural to expect the very early universe to be holographic and effectively $1+1$ dimensional. When the universe runs from the holographic stage to the radiation-dominated stage with normal entropy described by ordinary QFT, the equation of state (EoS) of the universe evolves from $w=1$ to $w=1/3$, and the effective dimension runs from $1+1$ to $3+1$.

The dimensional reduction at high energies in cosmology has already been reported by the analysis of the cosmic ray experiments \cite{ms}. Moreover, requiring an early $1+1$ dimensional stage naturally leads to a scale-invariant spectrum for cosmological perturbations even without inflation. See \cite{ac1,ds,bf1}.

\vspace{0.38cm}

\uppercase\expandafter{\romannumeral4}. \textbf{Conculsion}
\vspace{0.12cm}

We have provided a microscopic derivation to the holographic thermodynamics, which has profound implications for the understanding of black holes and the very early universe. We suggest that the fundamental physics behind black holes and the very early universe may be $1+1$ dimensional; and the physical concepts, i.e., maximum entropy, dimensional reduction and unconventional EoS $w=1$ may be closed correlated among each other.  We also collected and presented the supportive evidences from the recent progress in the literature of quantum gravity.


\vspace{0.38cm}

\begin{thebibliography}{99}

\bibitem{Sasakura:1999xp}
N.~Sasakura, \emph{An Uncertainty relation of space-time,}
Prog. Theor. Phys. \textbf{102}, 169-179 (1999)
[arXiv:hep-th/9903146 [hep-th]].

\bibitem{Brustein:1999md}
R.~Brustein and G.~Veneziano,
\emph{A Causal entropy bound,}
Phys. Rev. Lett. \textbf{84}, 5695-5698 (2000)
[arXiv:hep-th/9912055 [hep-th]].

\bibitem{mathur}
  A.~Masoumi and S.~D.~Mathur,
\emph{An equation of state in the limit of high densities,}
  Phys.\ Rev.\ D {\bf 90} (2014) no.8,  084052
  [arXiv:1406.5798 [hep-th]].



\bibitem{Xiao:2019pkr}
Y.~Xiao,
\emph{Microscopic derivation of the Bekenstein-Hawking entropy for Schwarzschild black holes,}
Phys. Rev. D \textbf{101}, no.4, 046020 (2020)
[arXiv:1910.10678 [gr-qc]].

\bibitem{Xiao:2020hdt}
Y.~Xiao,
\emph{Running of effective dimension and cosmological entropy in early universe,}
Eur. Phys. J. C \textbf{80}, no.12, 1154 (2020)
[arXiv:2005.01415 [gr-qc]].

\bibitem{laus}
O.~Lauscher and M.~Reuter,
\emph{Ultraviolet fixed point and generalized flow equation of quantum gravity,}
Phys. Rev. D \textbf{65}, 025013 (2001)
[arXiv:hep-th/0108040 [hep-th]].

\bibitem{amb}
  J.~Ambjorn, J.~Jurkiewicz and R.~Loll,
  \emph{Spectral dimension of the universe,}
  Phys.\ Rev.\ Lett.\  {\bf 95}, 171301 (2005)
  [hep-th/0505113].

\bibitem{horava}
  P.~Ho\v{r}ava,
\emph{Spectral Dimension of the Universe in Quantum Gravity at a Lifshitz Point,}
  Phys.\ Rev.\ Lett.\  {\bf 102}, 161301 (2009)
  [arXiv:0902.3657 [hep-th]].

\bibitem{carlip1}
S.~Carlip,
\emph{Dimension and Dimensional Reduction in Quantum Gravity,}
Class. Quant. Grav. \textbf{34}, no.19, 193001 (2017)
[arXiv:1705.05417 [gr-qc]].

\bibitem{carlip2}
  S.~Carlip,
\emph{Dimension and Dimensional Reduction in Quantum Gravity,}
  Universe {\bf 5}, 83 (2019)
  [arXiv:1904.04379 [gr-qc]].

  \bibitem{bema}
  J.~D.~Bekenstein and A.~E.~Mayo,
 \emph{Black holes are one-dimensional,}
  Gen.\ Rel.\ Grav.\  {\bf 33}, 2095 (2001)
  [gr-qc/0105055].


\bibitem{pol2}
  G.~T.~Horowitz and J.~Polchinski,
\emph{A Correspondence principle for black holes and strings,}
  Phys.\ Rev.\ D {\bf 55}, 6189 (1997)
  [hep-th/9612146].

  \bibitem{vene}
  T.~Damour and G.~Veneziano,
\emph{Selfgravitating fundamental strings and black holes,}
  Nucl.\ Phys.\ B {\bf 568}, 93 (2000)
  [hep-th/9907030].

\bibitem{Brustein}
R.~Brustein and A.~J.~M.~Medved,
\emph{Black holes as collapsed polymers,}
Fortsch. Phys. \textbf{65}, no.1, 1600114 (2017)
[arXiv:1602.07706 [hep-th]].

\bibitem{ms}
J.~R.~Mureika and D.~Stojkovic,
\emph{Detecting Vanishing Dimensions Via Primordial Gravitational Wave Astronomy,}
Phys. Rev. Lett. \textbf{106}, 101101 (2011)
[arXiv:1102.3434 [gr-qc]].

\bibitem{ac1}
G.~Amelino-Camelia, M.~Arzano, G.~Gubitosi and J.~Magueijo,
\emph{Dimensional reduction in the sky,}
Phys. Rev. D \textbf{87}, no.12, 123532 (2013)
[arXiv:1305.3153 [gr-qc]].

\bibitem{ds}
D.~Stojkovic,
\emph{Vanishing dimensions,}
Mod. Phys. Lett. A \textbf{28}, 1330034 (2013)
[arXiv:1406.2696 [gr-qc]].

 \bibitem{bf1}
  T.~Banks and W.~Fischler,
 \emph{ Holographic cosmology 3.0,}
  Phys.\ Scripta T {\bf 117}, 56 (2005)
  [hep-th/0310288].

\end{thebibliography}
\end{document}